\documentclass[%
reprint,                     
amsmath,amssymb,
aps,
]{revtex4-2}

\usepackage{graphicx}
\usepackage{dcolumn}
\usepackage{bm}

\usepackage{times}
\usepackage[colorlinks=true, citecolor=blue, linkcolor=blue, urlcolor=blue]{hyperref}
\usepackage{comment}
\usepackage[normalem]{ulem}
\begin{document}
	
	\preprint{APS/123-QED}
	
	\title{ Dark-Mode Control of Contrasting Entanglement and Bell Nonlocality between Mechanical Oscillators}
		
	
	\author{Souvik Agasti}
	\email{souvik.agasti@uhasselt.be}
	
	\affiliation{
		IMOMEC division, IMEC, Wetenschapspark 1, B-3590 Diepenbeek, Belgium
	}%
	\affiliation{
		Institute for Materials Research (IMO), Hasselt University,	Wetenschapspark 1, B-3590 Diepenbeek, Belgium
	}%
	
	\author{Philippe Djorw\'e}
	\email{djorwepp@gmail.com}
	\affiliation{Department of Physics, Faculty of Science, 
		University of Ngaoundere, P.O. Box 454, Ngaoundere, Cameroon}
	\affiliation{Stellenbosch Institute for Advanced Study (STIAS), Wallenberg Research Centre at Stellenbosch University, Stellenbosch 7600, South Africa}
	
	\author{Xin Zhou}
	\email{xin.zhou@cnrs.fr}
	\affiliation{ CNRS, Universit\'e de Lille, Centrale Lille, Universit\'e Polytechnique Hauts-de-France, UMR8520, IEMN, Av.
		Henri Poincare, Villeneuve d’Ascq F-59650, France}
	
	
	\begin{abstract}

This study presents a detailed proposal for an optomechanical system consisting of two mechanical oscillators coupled to a common cavity, aimed at generating  pure and entangled two-mode squeezed mechanical steady states. We found that optomechanical coupling imperfections can enhance mechanical entanglement while simultaneously suppressing Bell nonlocality by reducing the purity of the mechanical state. This implies that Bell inequality violation may not occur where entanglement is maximal; instead, nonlocality can be observed in states with lower entanglement. To mitigate this trade-off, we introduce phase-dependent phonon hopping between the mechanical oscillators and show that Bell nonlocality can be selectively enhanced in specific dark-mode configurations, even when the overall entanglement is reduced. This contrasting behavior is attributed to changes in state purity associated with an imbalance in the occupations of the Bogoliubov modes. Compatible with existing microwave cavity optomechanical platforms, the proposed optomechanical scheme provides an experimentally accessible route for controlling nonlocal quantum correlations in multimode mechanical systems. It could serve as an attractive platform for the deployment of continuous-variable teleportation and high-fidelity quantum communication.
       
	\end{abstract}
	
	\maketitle
	

	\section{Introduction}\label{Introduction}
	
	Continuous-variable (CV) entangled Gaussian states have attracted significant interest owing to their wide range of applications in precision measurements \cite{precision_measurement1, precision_measurement2}, quantum communication \cite{quantum_communication1, quantum_communication2, quantum_communication3}, and high-fidelity quantum state transfer \cite{quantum_teleportation, Clerk_Fidelity, vitali_TMS_generator}. Among bipartite Gaussian CV states, the two-mode squeezed (TMS) state constitutes one of the most prominent examples. The entanglement between them is of both fundamental and practical importance, with applications extending from quantum memories \cite{Quantum_memory} to gravitational-wave metrology \cite{my_BAE}. Such entanglement naturally motivates investigations of hidden-variable theories and the limits of local realism through Bell inequalities \cite{mypaper_TMSV_filter_STD, mypaper_TMSV_filter}. Quantum nonlocality for spatially separated systems was originally formulated by Clauser, Horne, Shimony, and Holt (CHSH) \cite{bell_CHSH, bell_CH}, and subsequently extended to phase-space formulations by Banaszek and W'odkiewicz using Wigner quasiprobability distributions \cite{Banaszek_TMSV_bell, Banaszek_TMSV_bell_PRL}. These approaches establish criteria for detecting nonlocality in bipartite Gaussian CV states and emphasize the crucial role of state mixedness \cite{agasti_nonlocally_CV}, consistent with the observation that all nonlocally realizable bipartite Gaussian states must necessarily be entangled.\\
	
	In recent years, multimode optomechanical systems have attracted considerable interest because they provide a promising platform for generating nonclassical states and probing the quantum nature of massive mechanical degrees of freedom \cite{Equivalence_optomechanics_Kerr,Squeezing_mechanics_Clerk, BAE_clerk_measurements, dark_mode_2_mechanics, SQZ_experiment,Reservoir_Engineering_two_mode, Clerk_Fidelity, my_BAE}. The mechanical position and momentum quadratures are continuous variables, such systems provide a natural setting for studying CV quantum correlations, including squeezing and entanglement. A particularly relevant configuration consists of two mechanical modes coupled to a common cavity field, which mediates interactions between the modes and can act as an engineered dissipative reservoir. Within this configuration, the generation of pure entangled steady states has been proposed through engineering a single reservoir  \cite{T_m_s_mechanics_Clerk}. Moreover, when the system is driven by multiple tones, interference between the optomechanical interactions can be exploited to manipulate dark modes, thereby controlling generations of the mechanical mode squeezing and the entanglement \cite{dark_mode_2_mechanics,dark_mode_1_mechanics, Nori_suggestion_2}. Such schemes also enable squeezing of collective quadratures below the standard quantum limit (SQL) and cooling of Bogoliubov modes \cite{Squeezing_mechanics_Clerk, BAE_clerk_measurements, dark_mode_2_mechanics, SQZ_experiment, cooling_BG_modes_Mika, Nori_suggestion_Nonreciprocal_ground_state_cooling}. These developments are closely connected to the fact that TMS states can be naturally represented in terms of nonlocal bosonic Bogoliubov modes.\\
    
    

    For entangled mechanical oscillators, Bell nonlocality provides a particularly stringent means of certifying quantum correlations in massive degrees of freedom comprising billions of atoms. It is therefore of fundamental interest for investigating the quantum-to-classical transition \cite{zurek2003decoherence}. More broadly, nonclassical mechanical correlations, including squeezing and entanglement, may enable applications in force sensing \cite{precision_measurement2}, gravitational-wave detection \cite{my_BAE}, and quantum-state transduction across distributed quantum networks \cite{entanglement_micromechanical_oscillators}. These considerations motivate a systematic investigation of the relationship between mechanical entanglement and Bell nonlocality in optomechanical systems. Although entanglement is generally regarded as a prerequisite for nonlocal quantum correlations, entanglement and nonlocality represent distinct physical resources and may even exhibit inverse behavior in two-qubit systems \cite{inverse_nonlocality_entanglement}. It is therefore of considerable interest to investigate whether entanglement and Bell nonlocality can exhibit contrasting or even opposite responses to system parameters in optomechanical CV systems. \\

    In this work, we investigate CHSH Bell nonlocality between two
mechanical oscillators coupled to a common cavity and clarify its
relation to their steady-state entanglement.  Different from previous work \cite{li2017einstein}, we show that
optomechanical coupling imperfections can increase the logarithmic
negativity while simultaneously suppressing the Bell violation. We
trace this contrasting behavior to a reduction in state purity,
associated with an imbalance between the occupations of the two
Bogoliubov modes. We further demonstrate that phase-dependent phonon
hopping can partially compensate for this imbalance in selected
dark-mode configurations. As a result, Bell nonlocality can be
enhanced even though the overall mechanical entanglement is reduced.
These results establish a controllable nonmonotonic relation between
entanglement and Bell nonlocality in a Gaussian optomechanical steady
state and identify state purity as the physical quantity governing
their contrasting responses. In addition, our proposed configuration provides a natural platform for exploring collective quantum dynamics of mechanical oscillators and constitutes a basic building block toward scalable multimode optomechanical architectures \cite{chegnizadeh2024quantum,massel2012multimode}. \\
     
     The rest of the paper is organized as follows. In Sec.~II, we introduce the Hamiltonian of a three-mode closed-loop optomechanical system, describe the reservoir-engineering protocol used to generate entanglement between two mechanical modes, derive the corresponding Quantum Langevin Equations (QLEs), and present their solutions. In Sec.~III, we obtain measures of entanglement and CHSH nonlocality from the correlations between the two mechanical oscillators and analyze their interplay. In particular, we examine the influence of dark modes on the inverse relationship between entanglement and nonlocality. In Sec.~IV, we discuss possible experimental implementations of the proposed scheme, and Sec.~V concludes our work.
	
    	\begin{figure}
		\includegraphics[width= 1 \linewidth]{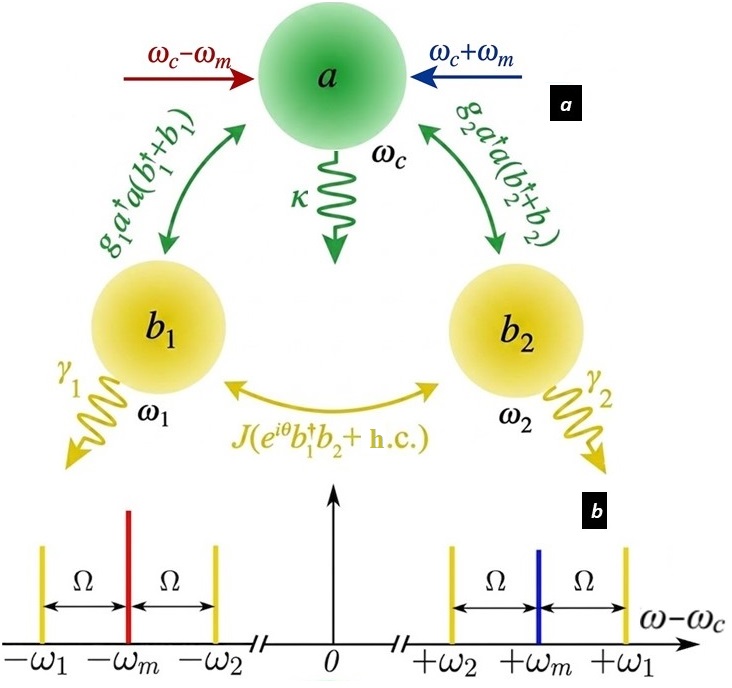}
		\caption{ (a) Block diagram of the single-cavity double-drum optomechanical setup, driven by a red and a blue drive. (b) The driving conditions, in terms of frequencies. The cavity resonance frequency is in the center, while the corresponding driving tones are indicated by blue and red lines, representing their role. The mechanical oscillation frequencies, indicated by the yellow lines, are placed symmetrically (at $\omega_1, \omega_{2}$) about detunings of $\Omega$ around the average frequency $\omega_m$. 
		}\label{block_diagram_dark_mode}
	\end{figure}
  
	\section{Optomechanical System under Reservoir Engineering}
	
	\subsection{Two-tone drive Hamiltonian }
	
	We consider an optomechanical system made of two mechanical oscillators coupled to a common electromagnetic field, as sketched in Fig.~\ref{block_diagram_dark_mode}(a). The Hamiltonian of our proposed system reads (with $\hbar=1$),
	\begin{equation}
		H = H_{OM}+H_{MM}+H_{\mathrm{drive}}+H_{\mathrm{diss}},
	\end{equation}
	where the optomechanical interaction is described by,
	
    \begin{equation}\label{system_Hamiltonian}
		H_{OM}=\omega_c a^\dagger a+\sum_{k=1,2}\omega_k b_k^\dagger b_k
		+\sum_{k=1,2}g_k a^\dagger a (b_k+b_k^\dagger).
	\end{equation}
    
	Here, $a$ ($a^\dagger$) denotes the annihilation (creation) operator of the cavity mode, while $b_k$ ($b_k^\dagger$) stands for the annihilation (creation) operator  of the $k$th mechanical oscillator. The resonance frequencies of the cavity and mechanical modes are given by $\omega_c$ and $\omega_k$, respectively. Dissipations are phenomenologically incorporated through $H_{\mathrm{diss}}$, with cavity decay rate $\kappa$ and mechanical damping rates $\gamma_k$. The term $H_{MM}$ describes the phase-dependent phonon-exchange interaction between the two mechanical resonators. The single-photon optomechanical coupling strengths are defined as
	$
	g_k=\frac{\omega_c}{L}\sqrt{\frac{\hbar/2}{m_k\omega_k}},
	$
	where $\hbar$ is the Plank constant, $L$ is the effective cavity length and $m_k$ is the effective mass of the $k$th mechanical oscillator. To generate a hybrid Bogoliubov modes between mechanical oscillators, it is required the cavity to be driven by two coherent tones at frequencies $\omega_c\pm\omega_m$ (at the blue and red sideband of the cavity, respectively), resulting in the driving Hamiltonian,
	\begin{equation}\label{driving_Hamiltonian}
		H_{\mathrm{drive}}=
		\left(
		E_+ e^{+i\omega_m t}
		+
		E_- e^{-i\omega_m t}
		\right)
		e^{i\omega_c t}a
		+\mathrm{h.c.},
	\end{equation}
	where $E_\pm$ are the drive amplitudes of the blue and red pumps, respectively. The parameters 
	\begin{equation}
		\Omega=\frac{\omega_1-\omega_2}{2},
		\qquad
		\omega_m=\frac{\omega_1+\omega_2}{2},
	\end{equation}
	correspond to the difference and average mechanical frequencies, respectively. A schematic representation of the driving configuration is shown in Fig.~\ref{block_diagram_dark_mode}(b), where the two drives determine the effective mechanical detunings. To engineer a two-mode squeezed Bogoliubov state between the mechanical oscillators with effective oscillation frequency $\Omega\gg\gamma_{1,2}$, we move to a rotating frame defined by,
	\begin{equation}
		H_{\mathrm{rotate}}
		=\omega_c a^\dagger a
		+(\omega_1-\Omega)b_1^\dagger b_1
		+(\omega_2+\Omega)b_2^\dagger b_2.
	\end{equation}
    Transforming into the rotating frame and neglecting nonresonant contributions within the resolved-sideband regime $(\omega_m\gg\kappa)$, the linearized Hamiltonian becomes,

	\begin{align}\label{linearized_Hamiltonian}
		H_{OM} = \Omega (b_1^\dagger b_1 - b_2^\dagger b_2)  & + G_+ [(b_1 +b_2)a + h.c. ] \\
		& + G_- [(b_1 +b_2)a^\dagger + h.c. ]. \nonumber
	\end{align}
	
	The single photon optomechanical couplings are considered equal, and the effective optomechanical coupling strengths are given by,
	\begin{equation}\label{OM_coupling}
		G_\pm =\alpha_\pm  g_1 = \alpha_\pm g_2, \,\,
		\text{
			with}  \,\,
		\alpha_\pm=\frac{iE_\pm}{\pm i\omega_m-\kappa},
	\end{equation}

	which may be taken as real, as we can tune the phase of drives and we are working in the resolved sideband regime $\omega_1, \omega_2 \gg \kappa$. This also discards time-dependent contributions to Hamiltonian at Eq. \eqref{linearized_Hamiltonian} \cite{T_m_s_mechanics_Clerk}. The stability of the system is analyzed in Appendix~\ref{Stability_System} using the Routh--Hurwitz criterion. Eq.~\eqref{linearized_Hamiltonian} can be conveniently rewritten in terms of hybrid mechanical Bogoliubov modes as,

	\begin{align}\label{Bogoliubov_Hamiltonian}
		H_{OM} = \Omega (\beta_1^\dagger \beta_1 - \beta_2^\dagger \beta_2) + \mathcal{G} [(\beta_1^\dagger + \beta_2^\dagger)a + h.c. ], 
	\end{align}
	
	where the Bogoliubov operators are defined by,
	
	\begin{subequations}\label{bg_modes}
		\begin{align}
			\beta_1 (\Omega) &= b_1 (\Omega) \cosh(r) + b_2^\dagger (\Omega) \sinh(r), \\
			\beta_2 (-\Omega) &= b_2 (-\Omega) \cosh(r) + b_1^\dagger (-\Omega) \sinh(r). 
		\end{align}
	\end{subequations}
	
	These modes correspond to collective mechanical excitations at frequencies $\pm\Omega$ in the rotating frame. The effective coupling strength and squeezing parameter are determined through,
	
	\begin{equation} 
    \mathcal{G}=\sqrt{G_-^2-G_+^2}, \qquad \tanh(r)=\frac{G_+}{G_-}. 
	\end{equation} 
	In the special case $G_+=G_-$, Eq.~\eqref{Bogoliubov_Hamiltonian} reduces to the two-mode backaction-evading measurement scheme introduced in Ref.~\cite{BAE_clerk_measurements}, which has also found applications in quantum-noise suppression for gravitational-wave detectors \cite{my_BAE}. For $G_+\neq G_-$, the backaction-evading condition is relaxed, enabling the generation of two-mode squeezing between the mechanical oscillators. Furthermore, the parameter $\Omega$ may alternatively be tuned using a four-tone driving protocol, as discussed in Ref.~\cite{T_m_s_mechanics_Clerk}. In that case, the effective optomechanical couplings are modified, and an additional balancing condition must be satisfied. For non-degenerated single-photon optomechanical couplings $(g_1\neq g_2)$, the two-tone driving scheme no longer perfectly matches the sideband processes associated with the two oscillators in Eq.~\eqref{linearized_Hamiltonian}. This mismatch introduces an additional contribution to the Hamiltonian of the form,
	\begin{equation}\label{additional_Hamiltonian} H_{OM}^m = \left[ G_+^m(b_1^\dagger-b_2^\dagger) - G_-^m(b_1-b_2) \right] a^\dagger +\mathrm{h.c.}, 
	\end{equation} 
    
	where the imperfect coupling amplitudes are given by, 
    
    \begin{equation} 
    G_\pm^m = \pm {\alpha_\pm(g_1-g_2)}/{2},
    \end{equation} 

    and the effective optomechanical coupling in Eq. \eqref{OM_coupling} modifies to $G_\pm =\alpha_\pm  (g_1 + g_2)/2$. Such coupling imperfections significantly perturb the Hamiltonian of Eq.~\eqref{Bogoliubov_Hamiltonian} in the form of Bogoliubov modes defined by Eq.~\eqref{bg_modes}. As a consequence, the correlations between the two mechanical oscillators are degraded, leading to a suppression of Bell nonlocality, as it will be analyzed in the following sections.\\

	\subsection{Mechanical Dark-Mode}
	    
    In a multimode optomechanical system driven by multiple tones, the interplay between photon-phonon interactions induce interference effects that lead to the emergence of bright and dark modes, i.e., one mode becomes strongly coupled to the cavity field (bright mode), while the other becomes decoupled or weakly coupled (dark mode) \cite{dark_mode_experiment_1, Nori_suggestion_2}. Dark mode plays important role in optomechanical systems, as it protects against mechanical dissipation \cite{Optomechanical_Dark_Mode}, enables high-fidelity state transfer between mechanical resonators \cite{dark_mode_fidelity, Nori_suggestion_Nature_3, Nori_suggestion_4}, and facilitates efficient routing and switching of phonons associated with distinct vibrational modes \cite{Dark_mode_nature}. Both bipartite and tripartite entanglement in optomechanical system have been investigated through optical dark mode control \cite{nori_tripartite_entanglement_dark_mode}. 
	  To lift this dark-mode protection, in this work, we introduce a phase-dependent phonon-hopping interaction between the two mechanical resonators, characterized by coupling strength $J$ and phase $\theta$ \cite{dark_mode_2_mechanics},

	\begin{equation}
		H_{MM} = J \left(e^{i\theta}b_1^\dagger b_2 + h.c.\right).
	\end{equation}
	
	The phase $\theta$ represents the relative phase between the two mechanical modes and can be experimentally controlled through the relative phase between the pump tone and the demodulation reference signal.  
	 To study the dark mode effects, it is convenient to fix $\theta$ together with the photon tunneling interaction $J$. Recently, both theoretical \cite{dark_mode_theory1, dark_mode_theory2, Nori_suggestion_Nonreciprocal_ground_state_cooling} and experimental \cite{dark_mode_experiment_1, dark_mode_experiment_2, dark_mode_experiment_3} works demonstrate that such a phase-dependent synthetic gauge field can be introduced in a three-mode closed-loop optomechanical platform.

	The reduced Hamiltonian of Eq.\ref{linearized_Hamiltonian}, in the scheme of the non-degenerated single-photon optomechanical couplings, becomes 
	
	\begin{align}
		H &= \Omega( b_1^\dagger b_1 - b_2^\dagger b_2)  + J (b_1^\dagger b_2 e^{i\theta} + b_2^\dagger b_1 e^{-i\theta})\\
		&+ \sum_{k=1,2} G_{0k}(a^\dagger b_k + a b_k^\dagger) +\sum_{k=1,2} G_{1k}(a^\dagger b_k^\dagger + a b_k) +H_{diss},  \nonumber
	\end{align}
	where $G_{11} = G_+ + G_+^m, G_{12}= G_+ - G_+^m , G_{01} = G_- - G_-^m, G_{02}= G_- + G_-^m  $. The dark-mode associated with the absence of a synthetic gauge field $J=0$ is discussed in Appendix \ref{dark_mode_J=0}. To further analyze the interference effects induced by the synthetic phase $\theta$, we introduce hybridized mechanical modes,
	
	\begin{equation*}
		\begin{cases} 
			B_+ &= f b_1 - e^{i\theta} h b_2 \\ 
			B_- &= e^{-i\theta} h b_1 + f b_2
		\end{cases} \implies
		\begin{cases} 
			& b_1 = f B_+ + e^{i\theta} h B_- \\ 
			& b_2 = f B_- - e^{-i\theta} h B_+ 
		\end{cases},
	\end{equation*}
	with the normalization condition $f^2+h^2=1$. Substituting these expressions into the Hamiltonian yields,
	
	\begin{align}\label{dark_Hamiltonian}
		H &= \sum_{j=\pm} ( {\Delta}_j B_j^\dagger B_j + F_j^* a B_j^\dagger + F_j B_j a^\dagger) \\
		& + \sum_{j=\pm} ( \tilde{F}_j^* a^\dagger B_j^\dagger + \tilde{F}_j B_j a)  +H_{diss}, \nonumber
	\end{align}
	where we have defined $ \Delta_{\pm} = \pm  \sqrt{ \Omega^2 + J^2} $, $ F_+ = fG_{01} - e^{-i\theta}hG_{02}$, $F_- = e^{i\theta}hG_{01} + fG_{02}$, and $\tilde{F}_+ = fG_{11} - e^{-i\theta}h G_{12}, \tilde{F}_- = e^{-i\theta}h G_{11} + f  G_{12}$, with,
	
	\begin{equation}\label{dark_mode_papras}
		\begin{cases} 
			f &= \frac{|{\Delta}_- - \Omega |}{\sqrt{( {\Delta}_- - \Omega)^2 + J^2}} \\ 
			h &= \frac{J f}{ \Delta_- - \Omega } 
		\end{cases},
	\end{equation}
	
      From Eq. \eqref{dark_mode_papras}, $F_\pm$ and $\tilde{F}_\pm$ are determined as,
	
	\begin{subequations}
		\begin{align}
        F_\pm = fG_-Z_\pm, \,  Z_\pm= \left[(1\mp\frac{G^m_-}{G_-}) \pm \frac{J(1\pm\frac{G^m_-}{G_-})}{\Omega + \sqrt{\Omega^2 + J^2}} e^{\mp i\theta}\right]   \\
        \tilde{F}_\pm= fG_+ \tilde{Z}_\pm, \, \tilde{Z}_\pm= \left[(1\pm\frac{G^m_+}{G_+}) \pm \frac{J(1\mp\frac{G^m_+}{G_+})}{\Omega + \sqrt{\Omega^2 + J^2}} e^{\mp i\theta}\right]  
		\end{align}
	\end{subequations}	

For the angles $\theta=(0,\pi)$, one of $F_\pm$ and $\tilde{F}_\pm$ goes to its minimum value, decoupling the mechanical mode from the cavity mode, which appears to be a dark mode effect. This gives us freedom to engineer on the dark mode to control entanglement and nonlocality.  The engineering on dark-modes, without and with the presence of coupling imperfections, are furthermore discussed in Sec. \ref{sub_PhononHopping}.
	
	\subsection{Langevin's Equation of Motion}\label{Equation_of_Motion}

	The Heisenberg-Langevin equation of motion of the system is given by,
	
	\begin{equation}\label{EQM_time}
		\dot{\mathbf{u} }  = A \mathbf{u} +\mathbf{u}^{in},  
	\end{equation}
	
	where,	
	
	\begin{widetext}			
		\begin{align} \label{A_Matrix}
			A = { \left(
				\begin{array}{cccccc}
					-\gamma_1 & \Omega  & J \sin \theta & J \cos \theta &  0 & G_- - G_+ - G^m_s\\
					-\Omega  & -\gamma_1 & -J \cos \theta & J \sin \theta & -G_- -G_+ +G^m_d & 0 \\
					-J \sin \theta & J \cos \theta & -\gamma_2 & -\Omega  & 0 & G_- -G_+ +G^m_s \\
					-J \cos \theta & -J \sin \theta & \Omega  & -\gamma_2 & -G_- -G_+ -G^m_d & 0 \\
					0 & G_- -G_+ -G^m_s & 0 & G_- -G_+ +G^m_s & -\kappa & 0 \\
					-G_- - G_+ +G^m_d & 0 & -G_- - G_+  -G^m_d & 0 & 0 & -\kappa \\
				\end{array} 
				\right) }, 
		\end{align}
	\end{widetext}	
    
	with $G^m_d=G^m_-- G^m_+$ and  $G^m_s=G^m_-+ G^m_+$, and	$\mathbf{u}^T=[Q_1,P_1,Q_2,P_2,X,Y]$ where $Q_k = (b_k + b_k^\dagger)/\sqrt{2}, P_k = -i(b_k - b_k^\dagger)/\sqrt{2}$ and $X = (a + a^\dagger)/\sqrt{2}, Y = -i(a - a^\dagger)/\sqrt{2}$ are the amplitude and phase quadratures of cavity and $k$th mechanical oscillator, respectively. Similarly,  ${\mathbf{u}^{in}}^T=[\sqrt{2\gamma_1}Q_{1}^{in}, \sqrt{2\gamma_1}P_1^{in}, \sqrt{2\gamma_2}Q_2^{in}, \sqrt{2\gamma_2}P_2^{in}, \sqrt{2\kappa}X^{in},\sqrt{2\kappa}Y^{in}]$ where $Q_{k}^{in} = (b_k^{in} + {b^{in}_k}^\dagger)/\sqrt{2}, P_{k}^{in} = -i(b_k^{in} - {b_k^{in}}^\dagger)/\sqrt{2}$ and $X^{in} = (a^{in} + { a^{in}}^\dagger)/\sqrt{2}, Y^{in} = -i(a^{in} - {a^{in}}^\dagger)/\sqrt{2}$ are the amplitude and phase quadratures and $ a^{in}({a^{in}}^\dagger),  b_k^{in}({b_k^{in}}^\dagger)$ are the anihilation (creation) field operators of the inputs of cavity and $k^{th}$ mechanical oscillator, respectively. According to the Routh-Hurwitz stability criteria, the system reaches a stable steady state only when all eigenvalues of the drift matrix $A$ possess negative real parts (see Appendix~\ref{Stability_System}). The correlation function of the input noise to the cavity is given by,
	
	\begin{subequations}
		\begin{align}
			\langle  a^{in}(t)  {a^{in}}^\dagger (t') \rangle &= [N(\omega_{a}) +1] \delta(t-t'), \\
			\langle\ {b_k^{in}}^\dagger(t) b_k^{in}(t') \rangle &= n_{m_k} \delta(t-t'),
		\end{align}
	\end{subequations} 
    
	where $N(\omega_c) = (e^{\hbar \omega_c/k_BT}-1)^{-1}$ are the mean thermal occupation numbers of the cavity reservoir and $n_{m_k}= (e^{\hbar \omega_k/k_BT}-1)^{-1}$ are the mean thermal occupation numbers of the $k^{th}$ mechanical reservoir. Conveniently, here we consider $n_{m_1}=n_{m_2}=n_m$. As the cavity frequency is very high $\hbar \omega_c/k_BT \gg 1$, the thermal bath moreover behaves as a vacuum $(N(\omega_c) \approx 0)$.

	\subsection{Bipartite correlation: Entanglement and nonlocality}

	The steady state correlation matrix can be determined by solving the following Lyapunov equation,
	
	\begin{equation}
		AV +VA^T = -D,
	\end{equation}
	where $D = \langle \{\mathbf{u}^{in} , {\mathbf{u}^{in} }^T \} \rangle/2 = \text{diag}[(2n_m + 1) \gamma_1,(2n_m + 1)\gamma_2, \kappa, \kappa]$, and $V$ is the correlation matrix, of which elements are given by $V_{ij} = \langle \{\mathbf{u} , {\mathbf{u} }^T \} \rangle/2 = \frac{1}{2} \langle u_i u_j + u_j u_i \rangle$, where $1 \leq [i,j] \leq 6 $. 
	
	The steady state entanglement between two mechanical oscillators ($M1$ and $M2$) is calculated by means of logarithmic negativity \cite{Simon}, defined by,
	
	\begin{equation}\label{log_negv}
		E_n = \max[0, -\ln 2\eta^-],
	\end{equation}
	
	where,
	
	\begin{equation}\label{eta}
		\eta^- = \sqrt{\frac{ 1}{2} \left( \Sigma (V_M) - \sqrt{\Sigma (V_M)^2 - 4 \det (V_M)} \right) }.
	\end{equation}
	In Eq.~\eqref{eta},  $ \Sigma(V_M) = \det (V_M^{11}) + \det (V_M^{22}) - 2 \det (V_M^{12}) $, with 
	
	\begin{align}\label{define_Vmatt}
		V_M &= \left[ \begin{array}{c c}
			V_M^{11} & V_M^{12} \\
			V_M^{21} & V_M^{22}
		\end{array} \right], 
	\end{align}
where $V_M^{11}$ ($V_M^{22}$) characterizes the first (second) mechanical resonator, while $V_M^{12}$ captures the correlation between them.  This result implies that a Gaussian state is entangled $(E_N>0)$ if and only if $\eta^-<1/2$, which is equivalent to Simon’s necessary and sufficient positive-partial-transpose criterion for bipartite Gaussian states \cite{Simon}. In terms of the covariance matrix, this condition can be expressed as
	$4 \det (V_M) < \Sigma(V_M ) - \frac{1}{4}$. \\

	Bell nonlocality is quantified through the maximized CHSH Bell function. For bipartite continuous-variable Gaussian states, the Bell function can be evaluated using the Banaszek--W\'odkiewicz phase-space formulation based on the Wigner quasiprobability distribution, together with local linear Bogoliubov transformations \cite{agasti_nonlocally_CV}. For a general bipartite Gaussian state characterized by the covariance matrix $V_M$, the maximally optimized Bell function is given by,
	
	\begin{equation}\label{nonlocality_formula}
		|B|_m = \mu \left[1 + \left(\frac{ \sqrt{lm} }{ \sqrt{ lm } +\tilde{c} }\right)^{\frac{ \sqrt{ lm } }{ \sqrt{ lm } +2\tilde{c} }} \left( \frac{ \sqrt{ l m } +2\tilde{c} }{ \sqrt{ l m } +\tilde{c} }\right) \right],
	\end{equation}
	where  $\det(V_M^{11}) = l^2, \det(V_M^{22}) = m^2, \det(V_M^{12}) = c_1c_2$ and $\det(V_M) = (lm-c_1^2)(lm-c_2^2) $ and $\tilde{c} = \max[|c_1|,|c_2|]$. To measure the purity of the state, we use the following definition,
	
	\begin{equation}\label{mixedness}
		\mu = \frac{1/4}{ \sqrt{\det(V_M )} }.
	\end{equation}
	A bipartite Gaussian state exhibits Bell nonlocality when the CHSH inequality is violated, namely when $|B|_m>2$. Larger values of $B_{\max}$ correspond to stronger nonlocal correlations. For Gaussian states, the maximum attainable Bell violation is $|B|_m\simeq 2.19$ \cite{agasti_nonlocally_CV}. In general, entanglement provides a necessary and sufficient condition for nonlocality only in the case of pure states. For mixed states, entanglement remains necessary but is no longer sufficient for Bell nonlocality, whereas the observation of Bell nonlocality always certifies the presence of entanglement \cite{mypaper_TMSV_filter, agasti_nonlocally_CV}.

	\begin{figure*}
		\includegraphics[width= 1 \linewidth]{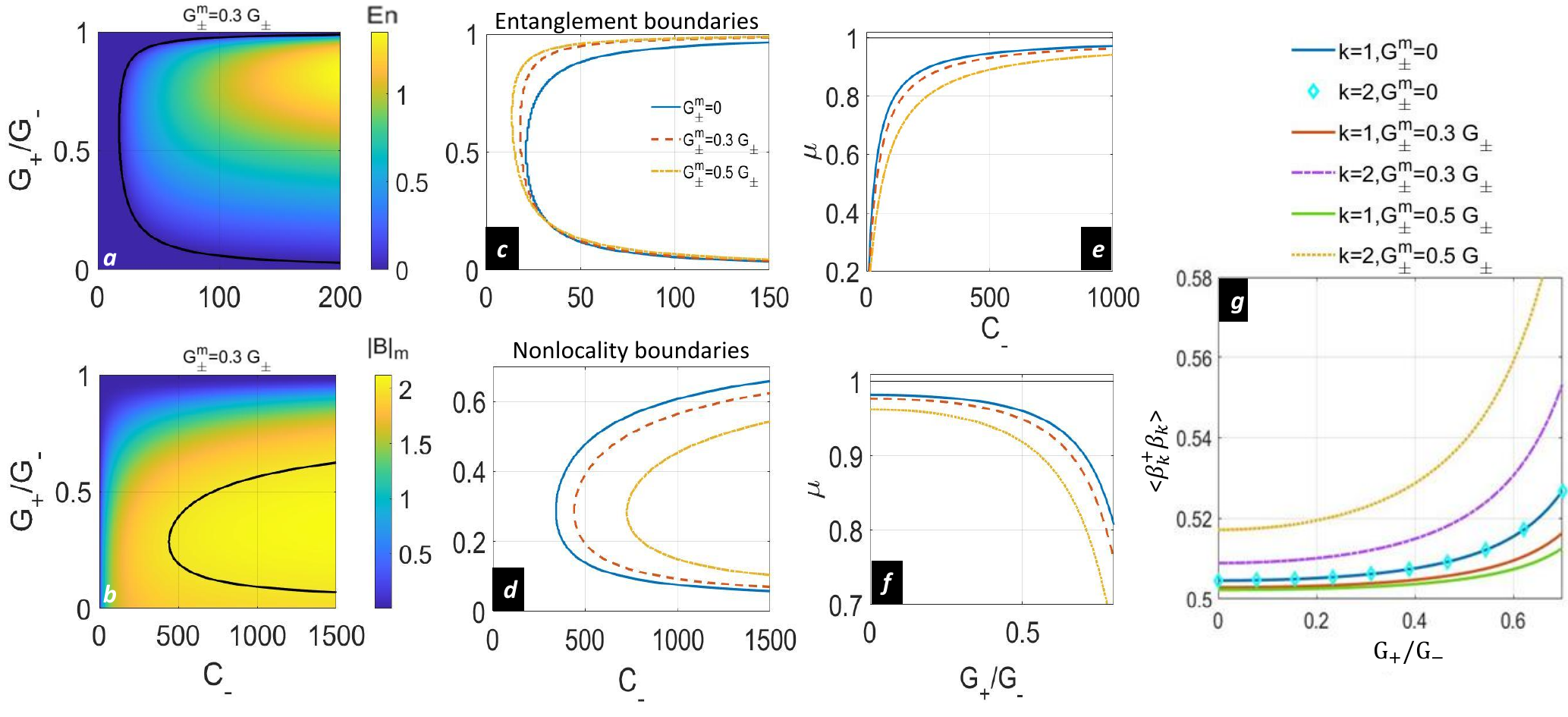}
		\caption{ (a) Entanglement measured by logarithmic negativity and (b) maximal value of Bell's function as a function of cooperativity ($C_- = 4G^2/\kappa\gamma_{1,2}$) and the coupling ratio $G_+/G_-$ for $G_\pm^m/G_\pm = 0.3$. Throughout this analysis, $\kappa$ and $\gamma_1=\gamma_2$ are kept fixed, while $G_-$ has been varied. The black line in (a) represents entanglement boundaries ($E_n = 0$), while in (b) it stands for the nonlocality boundary ($|B|_m = 2$). (c) Entanglement and (d) nonlocality boundaries for different coupling imperfections. In both (c) and (d), the region to the right of the boundary corresponds to the regime where entanglement and nonlocality exist between the two mechanical oscillators. (e) purity of the state ($\mu$) vs $C_-$ at $G_+/G_- = 0.3$, and (f) vs $G_+/G_-$ at $C_- = 1200$.  (g) Bogoliubov occupation number vs $G_+/G_-$ for different coupling imperfections at $C_- = 1200$.  In this case, the coupling between mechanical oscillators has been avoided $(J=0)$.
		Other parameters are, for example cavity linedidth $ \kappa = 2\pi\times 1.592 \times 10^5 \text{Hz} = 10^6 s^{-1} , \gamma_1 = \gamma_2 = 4 \times 10^{-5} \kappa, \Omega = 0.1\kappa $ and the corresponding thermal population $ n_m = 5$. This parameter set remains in the range of experimental parameters used in experiments conducted in \cite{Teufel_om_experiment}, which has also been followed in \cite{T_m_s_mechanics_Clerk}. 
		}\label{EN_NL}
	\end{figure*}
	
	\begin{figure}[t!]
		\includegraphics[width= 0.8 \linewidth]{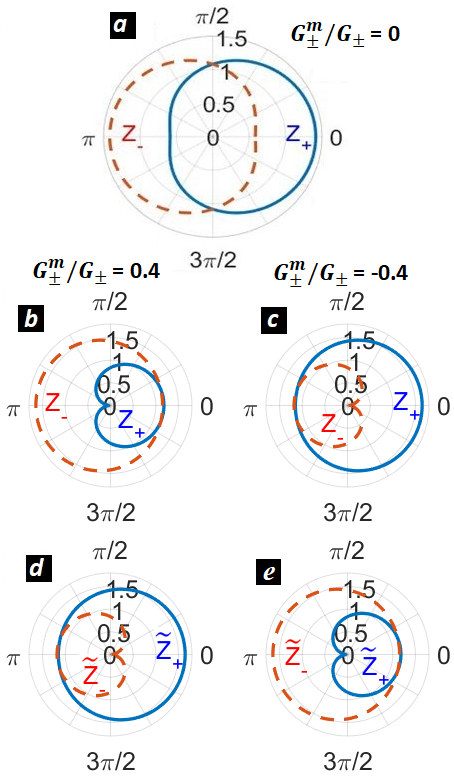}
		\caption{ (a) In case of absence of coupling imperfections $(G^m_\pm = 0)$ the redefined coupling factors $(Z_\pm = \tilde{Z}_\pm)$ are plotted versus the modulation phase $\theta$ in polar coordinates. The dark mode of the system appears at $\theta = (0,\pi)$. $\theta \neq (0,\pi)$ is the regime of dark mode breaking.
        In case of presence of coupling imperfections $(G^m \neq 0)$, $Z_\pm$ is plotted in (b,c) and $\tilde{Z}_\pm$ in (d,e). The used coupling imperfections are $ G^m_\pm/G_\pm = 0.4$ in left side (b,d) and $G^m_\pm/G_\pm = -0.4$ in right side (c,e). The parameter used here is $J= \Omega$ remains same for all the plots.
		}\label{DM_impurity}
	\end{figure}

	\begin{figure*}
		\includegraphics[width= 1 \linewidth]{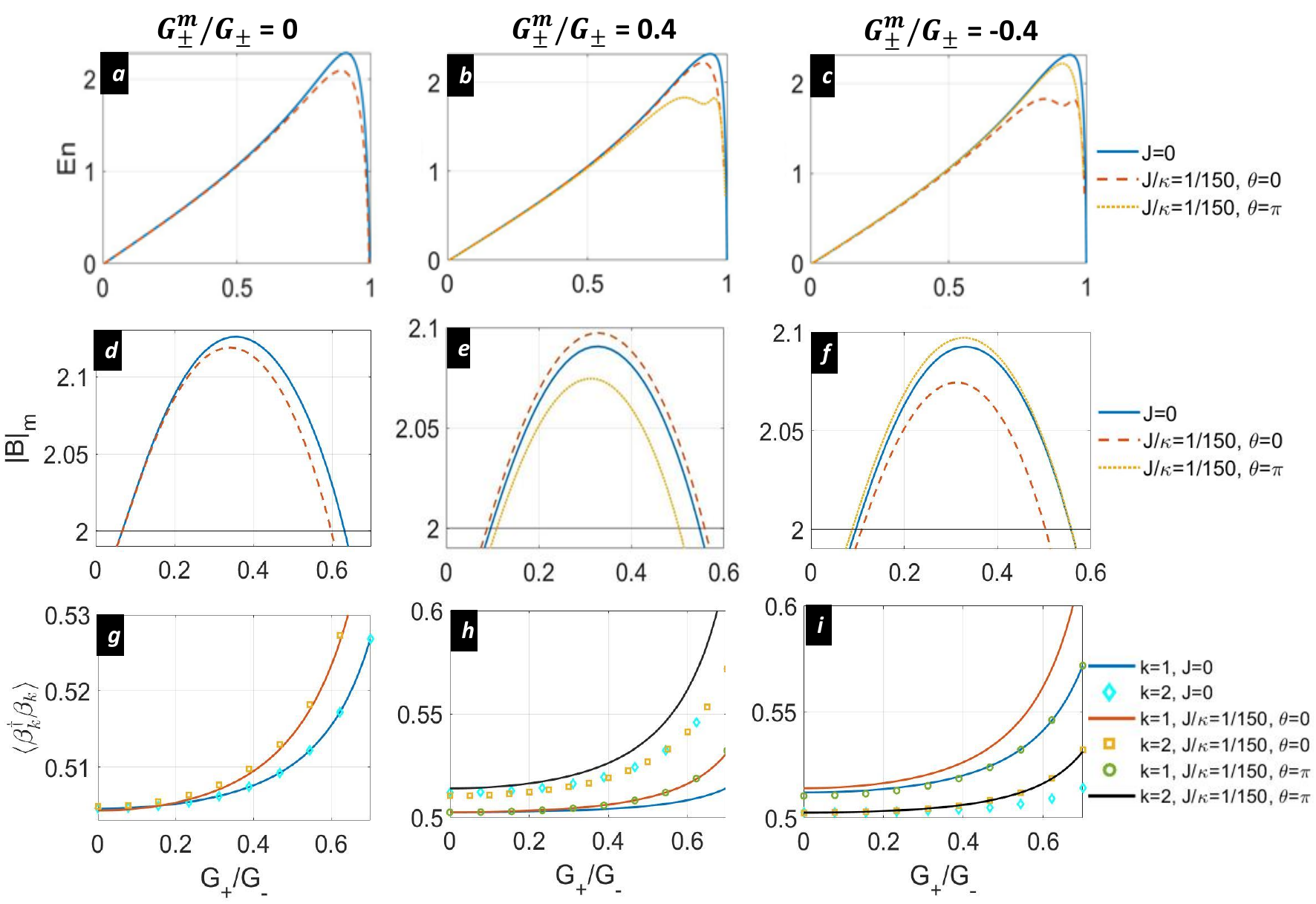}
		\caption{ Entanglement in top row (a,b,c), nonlocality in middle row (d,e,f) and Bogoliubov occupation numbers in bottom row (g,h,i) vs $G_+/G_-$ for $G_\pm^m/G_\pm = 0$ at left column (a,d,g), $G_\pm^m/G_\pm = 0.4$ at middle column (b,e,h) and $G_\pm^m/G_\pm = -0.4$ at right column (c,f,i). In the middle row, the black solid line is the nonlocality boundary $|B|_m = 2$. 
		In case of all (a,d,g) both the lines of $\theta = \pi$ and $\theta = 0$ fall on the same lines.
		All other parameters remain the same as in Fig. \ref{EN_NL}.      
		}\label{Ent_nl_J}
	\end{figure*}

	\section{Results: 
		entanglement vs nonLocality }

	\subsection{ Role of Coupling Imperfections in the Absence of Phonon hopping}
	
	We first evaluate the entanglement and nonlocality properties in the condition of neglecting phonon hopping between the mechanical oscillators ($J = 0$), through using the logarithmic negativity defined in Eq.~\eqref{log_negv} and the optimized Bell function introduced in Eq.~\eqref{nonlocality_formula}. The results are present in Fig.~\ref{EN_NL}.
    Although the considered setup has previously been demonstrated to efficiently generate entanglement between mechanical oscillators \cite{T_m_s_mechanics_Clerk}, violation of the CHSH inequality occurs only in the large-cooperativity regime, namely for $C_- = 4G_-^2/\gamma_k\kappa \gg 1$. Increasing the TMS parameter $r$, equivalently the ratio $G_+/G_-$, enhances the mechanical entanglement, as illustrated in Fig.~\ref{EN_NL}(a). In contrast, the Bell function decreases with increasing squeezing and eventually falls below the threshold for CHSH violation, exhibiting a Bell-shaped dependence [Fig.~\ref{EN_NL}(b)]. This behavior highlights the nontrivial relationship between entanglement and nonlocality, i.e.,  highly entangled states do not necessarily exhibit stronger Bell violations. Instead, CHSH nonlocality is observed predominantly for states characterized by comparatively lower the squeezing factor $(G_+/G_-)$.  This effect can be understood from the mixedness of the state, shown in Fig.~\ref{EN_NL}(e,f), which plays a crucial role in limiting Bell inequality violations.\\	
	
	An even more intriguing behavior emerges in the presence of imperfections in the single-photon optomechanical couplings. In agreement with previous studies \cite{T_m_s_mechanics_Clerk}, the mechanical entanglement increases with increasing coupling mismatch $(G_\pm^m \neq 0)$, thereby extending the entangled regime toward lower cooperativity values $C_-$ and enlarging the accessible parameter space [Fig.~\ref{EN_NL}(c)]. Remarkably, the opposite trend is observed for Bell nonlocality. As shown in Fig.~\ref{EN_NL}(d), the Bell function is suppressed in the presence of coupling imperfections, leading to a substantial reduction of the parameter region in which CHSH inequality violations can be observed.\\
	
	The contrasting behavior between entanglement and Bell nonlocality can be explained by analyzing the mixedness of the state, quantified through Eq.~\eqref{mixedness}. As presented in Fig.~\ref{EN_NL}(e), significant Bell violations occur only in the regime of large cooperativity $C_-$, and relatively small $G_+/G_-$ (Fig.~\ref{EN_NL}(f)), where the purity of the state remains comparatively high. A similar mechanism explains the effect of coupling imperfections $(G_\pm^m \neq 0)$: the resulting reduction in the purity of the state suppresses nonlocality, even though the degree of entanglement continues to increase [Fig.~\ref{EN_NL}(c,d)].\\	
	
	The reduction of purity can be further understood from the breakdown of the Hamiltonian in Eq.~\eqref{Bogoliubov_Hamiltonian}, expressed in terms of the Bogoliubov-mode structure of the hybrid mechanical modes introduced in Eq.~\eqref{bg_modes}. This effect becomes evident when examining the occupation numbers of the two Bogoliubov modes, $n_{\beta_k}=\langle \beta_k^\dagger \beta_k\rangle$. Any imbalances between these populations signal an increase in the mixedness of the state. As shown in Fig.~\ref{EN_NL}(g), the occupations of the two Bogoliubov modes $(k=1,2)$ remain identical in the absence of coupling imperfections $(G_\pm^m=0)$. However, when imperfections are introduced $(G_\pm^m\neq0)$, a pronounced population imbalance develops and grows with increasing $G_+/G_-$, which is consistent with the increase in mixedness displayed in Fig.~\ref{EN_NL}(f).\\	
	
	 In the following subsection \ref{sub_PhononHopping}, we consider the existence of the synthetic gauge field, i.e. finite hopping of phonons $(J\neq0)$, and investigate its influence on both entanglement and Bell nonlocality.
	
 \subsection{Engineering The Dark Mode}\label{sub_PhononHopping}

	In the absense of coupling imperfections ($G^m_\pm = 0$), the phase-dependent factors of the coupling strengths $(Z_\pm = \tilde{Z}_\pm)$ are plotted in Fig. \ref{DM_impurity}(a). Both coupling amplitudes reach their minima at $\theta=(0,\pi)$, corresponding to the $"+"$ and $"-"$ branches. At $\theta=(0,\pi)$ one among the terms of $F_\pm$ and one among $\tilde{F}_\pm$ of the Hamiltonian in Eq.\ref{dark_Hamiltonian} disappear, decoupling one of the mechanical modes from the cavity mode, having the Dark mode effects. In the ideal case $(G_\pm^m=0)$, maximal hybridization of the mechanical modes occurs for $\theta=\pm\pi/2$, providing the optimal condition for breaking the dark mode.\\
    
	In presence of coupling imperfection $G^m_\pm \neq 0$, $Z_\pm$ and $\tilde{Z}_\pm$ are plotted in Fig. \ref{DM_impurity}(b-e), where it can be observed that, even in case of coupling imperfections $G^m_\pm\neq 0$, both coupling amplitudes attain minima at $\theta=(0,\pi)$, with odd $n$ as it was observed in case of without of coupling imperfections. However, the overlaping of coupling factors ($Z_\pm$) moves away from $\theta=\pm\pi/2$, shifting the optimal phase away from $\theta= \pm\pi/2$. Existence of coupling imperfection ($G^m_\pm \neq 0$) and the coupling between mechanical modes impacts on the correlation between mechanical modes, and therefore entanglement and nonlocalities between them. \\   
	
	Figure~\ref{Ent_nl_J} illustrates the behavior of entanglement and Bell nonlocality in the presence of direct coupling between the mechanical oscillators. As shown in Fig.~\ref{Ent_nl_J}(a,b,c), the entanglement decreases whenever the intermechanical coupling is activated $(J>0)$, irrespective of whether optomechanical coupling imperfections are absent in Fig.~\ref{Ent_nl_J}(a) or present in Fig.~\ref{Ent_nl_J}(b,c) ($G_\pm^m>0$ or $<0$). In contrast, the behavior of the Bell function displayed in Fig.~\ref{Ent_nl_J}(d,e,f) reveals a qualitatively different trend. Unlike entanglement, the Bell violation can be enhanced for one of the two dark-mode configurations (Fig.~\ref{Ent_nl_J}(e) and (f)), while being suppressed for the other. The specific dark mode exhibiting enhanced Bell nonlocality depends on the sign of the optomechanical coupling imperfections, i.e.,  the enhancement occurs for $\theta=0$ when $G_\pm^m>0$, whereas it shifts to $\theta=\pi$ for $G_\pm^m<0$. \\
	
	The contrasting behavior between entanglement and nonlocality can be understood by analyzing the mixedness of the state through the Bogoliubov-structure. In Fig.~\ref{Ent_nl_J}(g,h,i), we plot the occupation numbers of the Bogoliubov modes
	both in the absence and the presence of intermechanical coupling. We find that, when optomechanical coupling imperfections are absent $(G_\pm^m=0)$, activating phonon tunneling between the oscillators hardly induces any asymmetry in the Bogoliubov-mode populations [Fig.~\ref{Ent_nl_J}(g)]. By contrast, in the presence of coupling imperfections $(G_\pm^m\neq0)$, a clear population imbalance emerges [Fig.~\ref{Ent_nl_J}(h) and Fig.~\ref{Ent_nl_J}(i)]. Remarkably, this imbalance is reduced in the dark-mode configurations of the $\theta=0$ when $G_\pm^m>0$, and $\theta=\pi$ when $G_\pm^m<0$. These are precisely the parameter regimes in which the Bell function is enhanced in Fig.~\ref{Ent_nl_J}(e,f), despite the simultaneous reduction of entanglement shown in Fig.~\ref{Ent_nl_J}(b,c). This behavior indicates a (partial) restoration of the purity of the state, which in turn favors the increment of Bell nonlocality. For the complementary dark-mode configurations, namely $\theta=\pi$ when $G_\pm^m>0$ and $\theta=0$ when $G_\pm^m<0$, the population imbalance between the two hybrid Bogoliubov modes increases. Consequently, both entanglement and Bell nonlocality are suppressed, as illustrated in Fig.~\ref{Ent_nl_J}(a,b,c) and Fig.~\ref{Ent_nl_J}(d,e,f).\\
		
		
	
\begin{figure*}[t!]
		\includegraphics[width= 0.91 \linewidth]{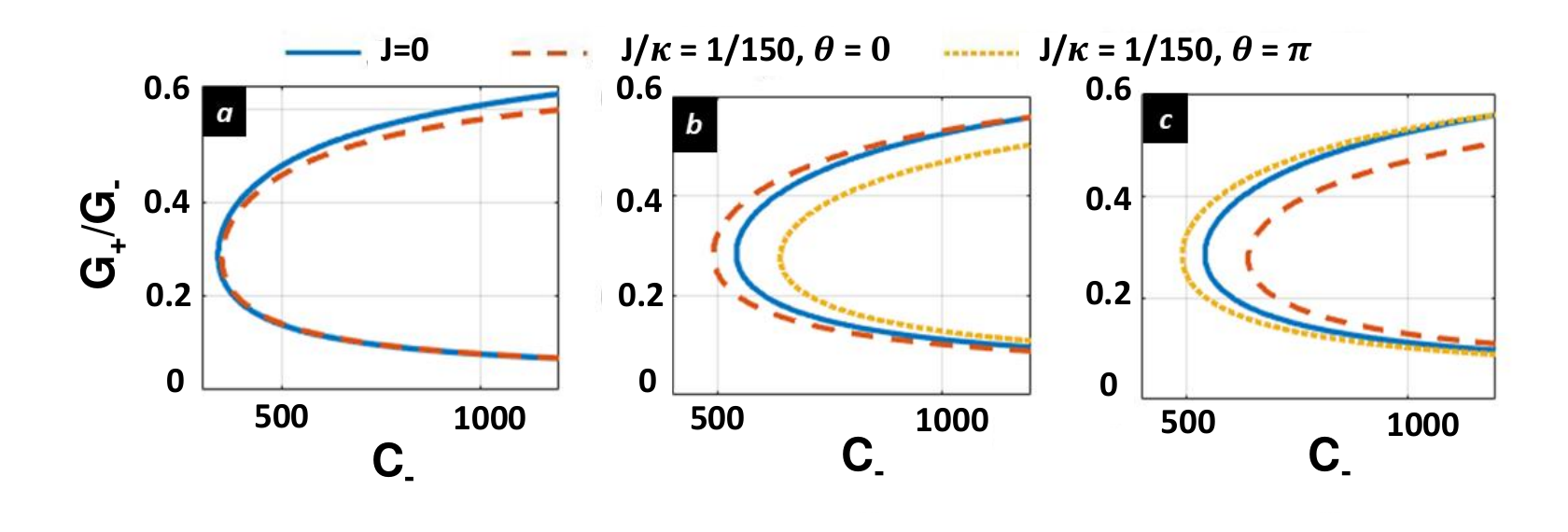}
		\caption{ Nonlocality boundaries $|B|_m = 2$ for different coupling imperfections (a) $G_\pm^m/G_\pm = 0$ (b) $G_\pm^m/G_\pm = 0.4$ and (c) $G_\pm^m/G_\pm = -0.4$. In all the cases, likewise (a) and (b) in Fig. \ref{EN_NL}, on the right side of the boundary, one can find nonlocality between two mechanical oscillators. 
			In case of (a), both the lines of $\theta = \pi$ and $\theta = 0$ fall on the same line.
			All other parameters remain the same as in Fig. \ref{EN_NL}.
		}\label{nl_boundary}
	\end{figure*}

	The modification of the CHSH Bell function under finite phonon hopping between the mechanical oscillators $(J>0)$ motivates an analysis of the parameter regimes in which nonlocal correlations can be observed. In Fig.~\ref{nl_boundary}, we present the nonlocality boundaries defined by the condition $B_{\max}=2$. The region to the right of each boundary corresponds to the parameter space where CHSH inequality violations occur, as illustrated previously in Fig.~\ref{EN_NL}(b).	
	Consistent with the suppression of the Bell function observed in Fig.~\ref{Ent_nl_J}(d), for all dark-mode configurations when no optomechanical coupling imperfections are present $(G_\pm^m=0)$,	the introduction of phonon hopping generally causes the nonlocality boundaries to contract [Fig.~\ref{nl_boundary}(a)]. In the presence of coupling imperfections, the contraction persists specifically for the dark modes characterized by $\theta=0$ when $G_\pm^m < 0$, and by $\theta=\pi$ when $G_\pm^m >0$.
	By contrast, for the complementary dark-mode configurations, namely $\theta=0$ with $G_\pm^m>0$ and $\theta=\pi$ with $G_\pm^m<0$, the nonlocality boundaries expand [Fig.~\ref{nl_boundary}(b and c)]. This behavior is consistent with the enhancement of the CHSH Bell function discussed previously and indicates an enlarged parameter regime supporting observable Bell nonlocality.\\

	\begin{figure*}[t!]
		\includegraphics[width= 1 \linewidth]{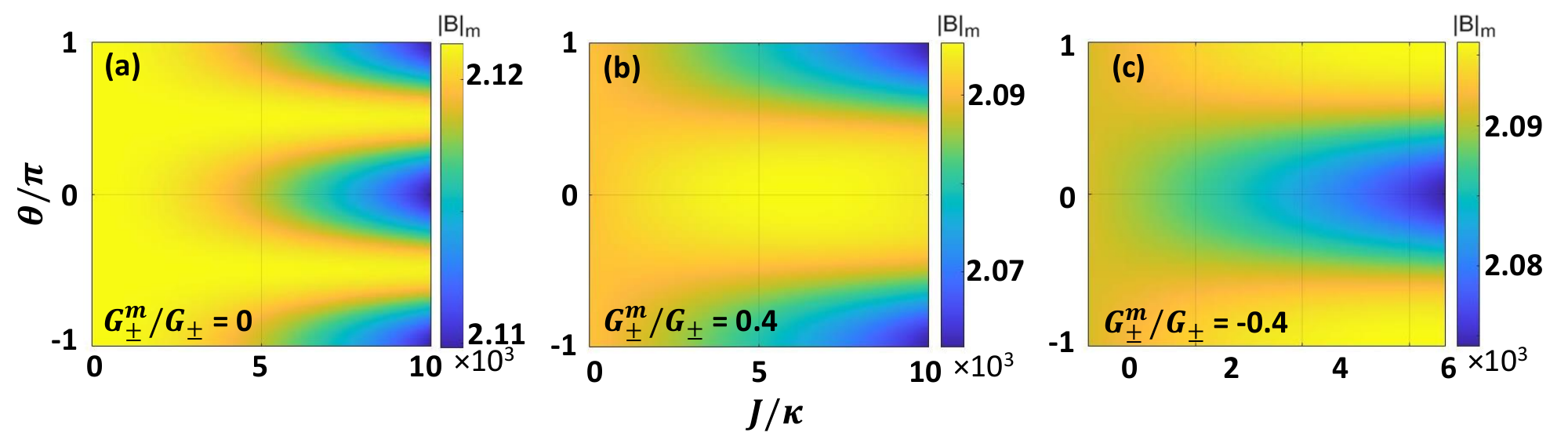}
		\caption{ $|B|_m$ for different coupling imperfections (a) $G_\pm^m/G_\pm = 0$ (corresponds to Fig. \ref{DM_impurity}(a), $Z_\pm$ are overlaping around $\theta=\pm\pi/2$ where $|B|_m$ maximizes) (b) $G_\pm^m/G_\pm = 0.4$ (corresponds to Fig. \ref{DM_impurity}(b and d), where the overlaping of $Z_\pm$ goes around $\theta=0$  where $|B|_m$ maximizes) and (c) $G_\pm^m/G_\pm = -0.4$ (corresponds to Fig. \ref{DM_impurity}(c and e), where the overlaping of $Z_\pm$ goes around $\theta=\pm\pi$ where $|B|_m$ maximizes). All other parameters remain the same as in Fig. \ref{EN_NL}.
		}\label{nl_J}
	\end{figure*}

	\begin{figure}[t!]
		\includegraphics[width= 1 \linewidth]{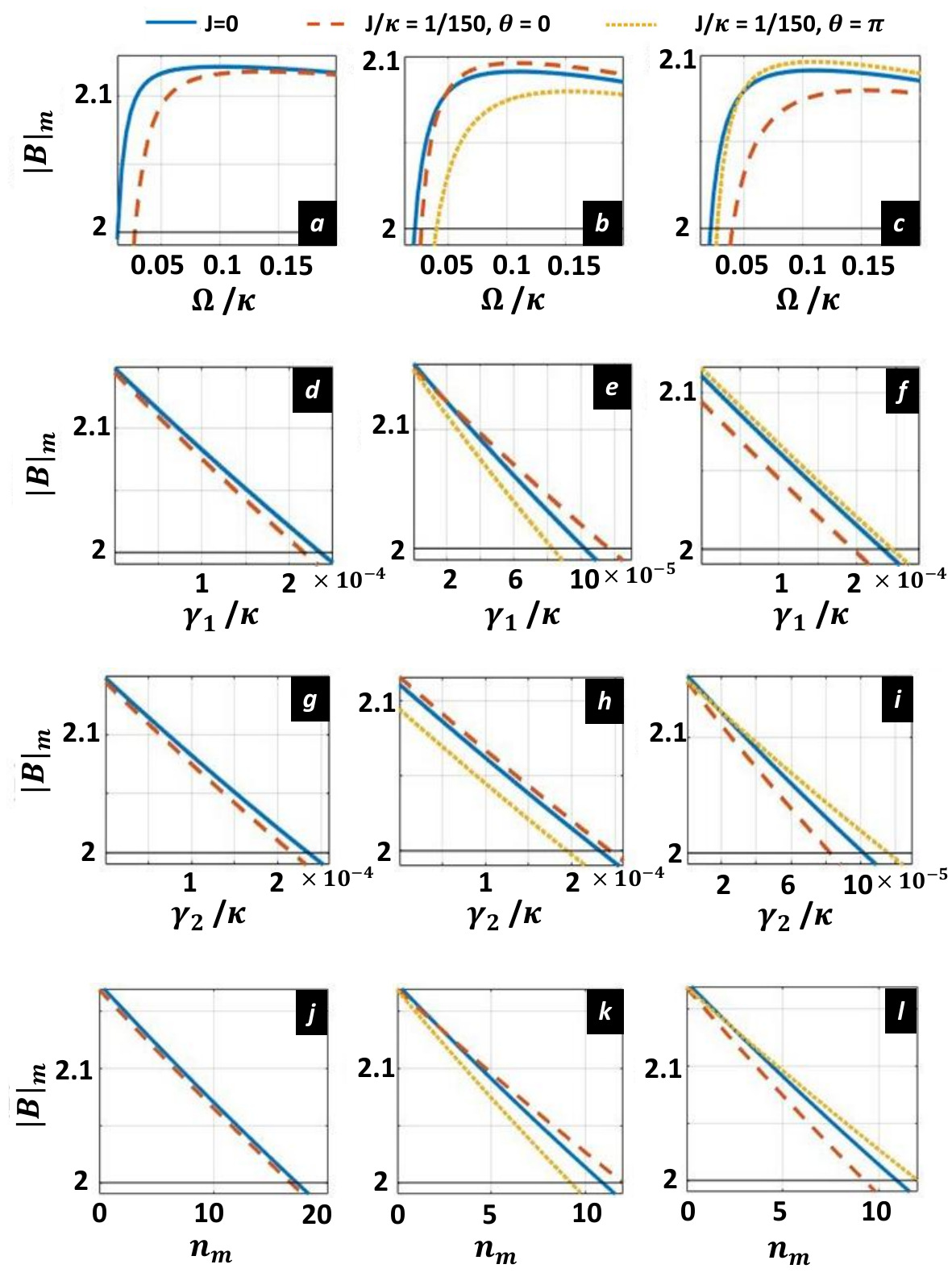}
		\caption{ $|B|_m$ vs $\Omega$ at the top row (a,b,c), $\gamma_1$ at the upper middle row (d,e,f), $\gamma_2$  at the lower middle row (g,h,i) and $n_m$ at the bottom row (j,k,l) 
			for different coupling imperfections: $G_\pm^m/G_\pm = 0$ at left column (a,d,g,j),  $G_\pm^m/G_\pm = 0.4$ at middle column (b,e,h,k), and $G_\pm^m/G_\pm = -0.4$ at right column (c,f,i,l). All the cases, nonlocality boundary (where $|B|_{max}= 2$ is indicated by the black solid line)
			In case of all the plots in left column (a,d,g,j) both the lines of $\theta = \pi$ and $\theta = 0$ falls on the same lines.
			All other parameters remain the same with Fig. \ref{EN_NL}.
		}\label{nl_all}
	\end{figure}

	\subsection{Bell Measurement Under Dark Mode Breaking }
	The enhancement of Bell nonlocality in specific mechanical dark-mode configurations motivates a more detailed investigation of the dependence of the CHSH Bell function on the phonon-hopping strength $J$ and, more generally, on deviations from the ideal dark-mode condition $(\theta \neq (0,\pi))$. Fig.~\ref{nl_J} presents the behavior of the optimized Bell function $|B|_{m}$ as a function of both $J$ and the phase parameter $\theta$. In the absence of optomechanical coupling imperfections $(G_\pm^m=0)$, the Bell function decreases uniformly with increasing phonon hopping and approaches reduced values near both configurations $\theta=(0,\pi)$ (Fig.~\ref{nl_J}(a)). In contrast, when coupling imperfections are present $(G_\pm^m\neq0)$, the behavior becomes strongly asymmetric. In this case, $|B|_{m}$ can be enhanced in one out of $0,\pi$ while being simultaneously suppressed in the other. Which among $0,\pi$ exhibits enhancement or suppression depends on the sign of the coupling imperfections, namely whether $G_\pm^m>0$ (Fig.~\ref{nl_J}(b)) or $G_\pm^m<0$ (Fig.~\ref{nl_J}(c)).\\

    With increasing phonon hopping, the maximum values of Bell's nonlocality is observed at the configurations where the modes are maximally hybridized, strongly coupled to the cavity mode, as indicated in Eq. \eqref{dark_Hamiltonian}. In the absence of optomechanical coupling imperfections $(G_\pm^m=0)$ (Fig.~\ref{nl_J}(a)), it appears at $\theta=\pi/2$, at which $Z_\pm$ and $\tilde{Z}_\pm$ observed overlaping in Fig. \ref{DM_impurity}(a). When the coupling imperfections are introduced $(G_\pm^m \neq 0)$ in Figs.~\ref{nl_J}(b and c), the maximum values of Bell's nonlocality moves away from $\theta=\pm\pi/2$, which remain consistant with the fact that the phase overlappings of $Z_\pm$ and $\tilde{Z}_\pm$ moves away from $\theta=\pm\pi/2$, which is explained in detail in Fig. \ref{DM_impurity}(b-e). The variation of Bell nonlocality under different coupling imperfection is determined in poler coordinate in Appendix. \ref{DM_coupling_imperfection}.  \\	
	
	\subsection{Role of Other System Parameters on nonlocal realization}
		

	We further investigate Bell nonlocality as a function of the mechanical frequency mismatch $(\Omega)$, as shown in Fig.~\ref{nl_all}(a,b,c). The analysis is performed both in the absence and presence of intermechanical phonon hopping and for the different dark-mode configurations. We find that the Bell function decreases rapidly as $\Omega$ becomes smaller for all considered cases. The suppression becomes even more pronounced when phonon hopping is activated $(J>0)$, independent of the specific dark mode. Moreover, the enhancement of nonlocality observed in particular $\theta=0$ for $G_\pm^m>0$ (Fig.~\ref{nl_all}(b)) and $\theta=\pi$ for $G_\pm^m<0$ (Fig.~\ref{nl_all}(c)), gradually disappears in the low $\Omega$ regime as the Bell function is strongly suppressed.\\
	
	The CHSH Bell function is also found to improve with increasing mechanical quality factors, $Q_j=\omega_j/\gamma_j$. This behavior is illustrated in Fig.~\ref{nl_all}(d-i), where $B_{\max}$ decreases with increasing one of the mechanical damping rates between $\gamma_{1,2}$, fixing the other unchanged. Physically, stronger mechanical dissipation weakens the correlations between the two oscillators, thereby reducing Bell nonlocality. The relative differences in the Bell's measurement ($|B|_{m}$) remains qualitatively unchanged (Fig.~\ref{nl_all}(f and h)) while varrying $\gamma_{1,2}$. In contrast, the differences in $|B|_{m}$ measurement between the two different dark-mode configurations become more prominent with the increment of $\gamma_{1,2}$, particularly for $G_\pm^m>0$ in Fig.~\ref{nl_all}(e), and for $G_\pm^m<0$, as shown in Fig.~\ref{nl_all}(i).\\
	
	Finally, we examine the influence of thermal decoherence on Bell nonlocality in the mechanical oscillators. As shown in Fig.~\ref{nl_all}(j,k,l), thermal dissipation does not qualitatively alter the observed behavior. In particular, the enhancement of nonlocality in the favored dark-mode configurations exhibits comparatively strong robustness against thermal decoherence (Fig.~\ref{nl_all}(k) and (l)).\\

	
	\section{Possible experimental realization}
		
	Throughout this work, we employ experimentally realistic parameters compatible with current microwave cavity optomechanical platforms. In particular, the chosen parameter regime is compatible with drumhead and double-drum micromechanical resonators \cite{Teufel_om_experiment,zhou2021high, pokharel2022coupling}, which can be easily integrated with conventional superconducting microwave cavities to form multimode optomechanical systems. The key difference from them 
    in the present proposal is the requirement of operating in a significantly higher cooperativity regime in order to observe Bell nonlocality. Nevertheless, all considered parameters remain within the validity regime of our theoretical treatment and are compatible with resolved-sideband operation.\\
	
	Recent experimental advances in cavity optomechanics have enabled the preparation of a variety of nonclassical mechanical states, including quantum ground states, squeezed mechanical states \cite{Squeezing_mechanics_Clerk, SQZ_experiment, dark_mode_2_mechanics}, and entangled states of nanomechanical resonators \cite{entanglement_micromechanical_oscillators}. In particular, such 3-mode closed-loop optomechanical circuits have been of fundamental interest for experimental realization of optomechanically induced non-reciprocal phase shift \cite{dark_mode_experiment_1, dark_mode_experiment_2, dark_mode_experiment_3}. These developments indicate that the implementation of the present scheme lies within reach of current experimental capabilities.\\
		
	\begin{figure}
		\includegraphics[width= 1 \linewidth]{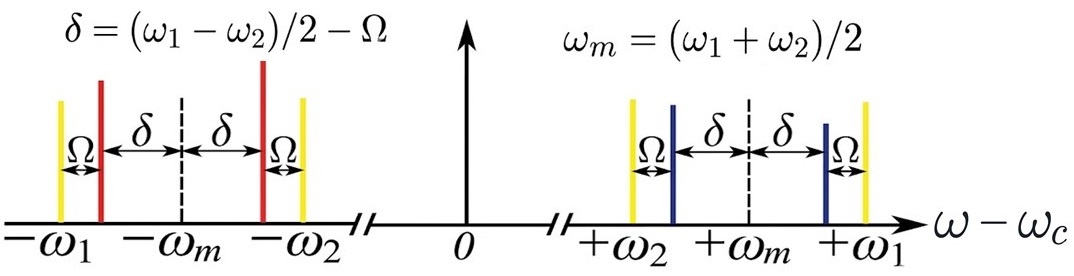}
		\caption{The 4-tone driving conditions, in terms of frequencies. The cavity resonance frequency is in the center, while the corresponding driving tones are indicated by blue and red lines, representing their role. The mechanical oscillation frequencies, indicated by the yellow lines, are placed symmetrically (at $\omega_1, \omega_{2}$) about detunings of $\Omega$ around the average frequency $\omega_m$. 
		}\label{freq_diagram_2}
	\end{figure}

	
	The Hamiltonian Eq. \eqref{linearized_Hamiltonian} involves four sideband processes; the up-conversion and down-conversion of drive photons via the absorption or emission of quanta from or to the mechanical oscillator 1 or 2. The realisation of Eq. \eqref{linearized_Hamiltonian} requires a balance of the rates at which these processes take place. By exploiting the four-tone driving scheme, one tone associated with	each sideband process, the balancing of the rates of these processes is possible even if the single-photon optomechanical couplings are unequal. In Fig. \ref{freq_diagram_2}, we show the four-tone driving field with frequencies $ \omega_c \pm (\omega_1 - \Omega)$ and $ \omega_c \pm (\omega_2 + \Omega)$ corresponding to the blue and red sidebands respectively. The frequency mismatch, in this case, modifies to

    \begin{equation}
        \Omega = \frac{\omega_1-\omega_2}{2} - \delta,
    \end{equation}

   where, $\delta$ is a shift of sidebands (both blue and red) from their center frequency. $\Omega$ enables effective frequency locking between the cavities, facilitating the formation of hybrid Bogoliubov modes. The driving Hamiltonian of Eq. \eqref{driving_Hamiltonian} becomes,
	
	\begin{align}
		H_{\mathrm{drive}} &= \big( E_{1+} e^{+i(\omega_1-\Omega) t} + E_{2+} e^{+i(\omega_2+\Omega) t} \\ 
		& + E_{1-} e^{-i(\omega_1-\Omega) t} + E_{2-} e^{-i(\omega_2+\Omega) t}\big) e^{+i\omega_c t} a + \mathrm{h.c.}, \nonumber
	\end{align}

	Furthermore, imposing a balancing condition,
	
	\begin{equation}\label{matching_condition}
		\frac{g_1}{g_2}= \frac{\alpha_{2 \pm }}{\alpha_{1 \pm }},
	\end{equation}
one obtains the many-photon optomechanical coupling rates (taken to be real),
	
	\begin{equation}\label{optomechanical_coupling}
		G_\pm =  (g_1 \alpha_{1 \pm } +  g_2 \alpha_{2 \pm })/2,
	\end{equation}
where the steady-state intracavity field amplitudes are denoted by	$\alpha_{k \pm } \equiv i E_{j \pm }/(\pm i\omega'_j - \kappa)$, with
	
	\begin{equation}
		\omega'_1 \equiv \omega_1 - \Omega \,\, \text{and} \,\, \omega'_2 \equiv \omega_2+ \Omega.
	\end{equation}
	
	The imperfection in the balancing condition in \eqref{matching_condition} produces additional Hamiltonian  $H^m_{OM}$ in  \eqref{additional_Hamiltonian}, with
	
	\begin{equation}
		G^m_\pm = \pm (g_1\alpha_{1 \pm } - g_2\alpha_{2 \pm })/2.
	\end{equation}
	
	The mechanical difference frequency shall be set in such a way that,
	
	\begin{equation}
		\gamma_{1,2} \ll \Omega \ll (\omega_1-\omega_{2})/2-\gamma_{1,2}.
	\end{equation}
	
	The rest of the process remains the same as it is done in the case of a 2-tone drive. However, the 4-tone drive gives extra freedom to tune over $\Omega$, and compensating single photon coupling mismatch.
	
	\section{CONCLUSION} 

    This work presents a theoretical framework for a three-mode optomechanical system in which two mechanical oscillators are coupled to a common cavity mode to generate highly pure and strongly entangled two-mode squeezed mechanical steady states. Our analysis reveals parameter regimes in which entanglement and Bell nonlocality exhibit contrasting, and even inverse, responses: optomechanical coupling imperfections can enhance mechanical entanglement while simultaneously suppressing Bell nonlocality through a reduction of the state purity. This behavior highlights that entanglement and Bell nonlocality, although closely related, constitute distinct physical resources and need  not to vary monotonically with the same system parameters.\\

    To control this trade-off, we introduce phase-dependent phonon hopping between the mechanical oscillators. In the presence of coupling imperfections, this mechanism enables selective enhancement of Bell nonlocality in specific dark-mode configurations, even when the overall entanglement is reduced. The enhancement remains robust in a finite range of thermal occupation and mechanical dissipation. We further identify state purity as the key physical quantity underlying the contrasting behavior of entanglement and nonlocality, with the imbalance of the Bogoliubov-mode occupations providing a physical signature of the increased mixedness of the mechanical state.\\
	
	The proposed three-mode optomechanical scheme is compatible with current experimental capabilities and therefore offers a promising platform for the realization of continuous-variable quantum teleportation and related quantum information protocols.

	\begin{acknowledgments}
		S.A. would like acknowledge the support of the European Commission, MSCA GA no 101065991 (SingletSQL). {P.D. is grateful to the Iso-Lomso Fellowship at Stellenbosch Institute for Advanced Study (STIAS), Wallenberg Research Centre at Stellenbosch University, Stellenbosch 7600, South Africa.}  X.Z. also would like to acknowledge financial support from the French National Research Agency, ANR-MORETOME, No. ANR-22-CE24-0020-01. {P.D. and X.Z. are  thankful to Campus France for the Partenariat Hubert Curien: PHC-BANTOU program 2026.}  
		
	\end{acknowledgments}

	\section*{ Disclosures }	
	The author declares no conflicts of interest.

	\appendix
	
	\section{Stability of the System} \label{Stability_System}
	
\begin{figure}[t!]
		\includegraphics[width= 1 \linewidth]{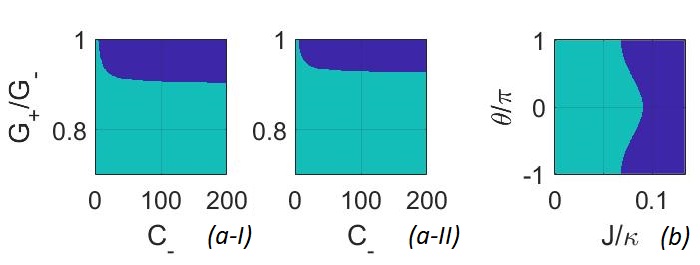}
		\caption{ Stability of the system for $G_\pm^m/G_\pm = \pm 0.4$ and (a-1) $J/\kappa = 1/15, \theta= 0$, (a-II)  $J/\kappa = 1/15, \theta= \pi$ and (b) $C_- = 1200, G_+/G_-=0.9$ All other parameters remain the same with Fig. \ref{EN_NL}.
		}\label{stability}
	\end{figure}

	The stability region as a function of the optomechanical couplings and the photon-hopping parameters  $(J, \theta)$ is shown in Fig. \ref{stability}. The figure confirms that the parameter regime of interest lies well within the stable region. System stability is determined using the Routh–Hurwitz stability criterion, which imposes the nontrivial condition that the real parts of all eigenvalues of the matrix $A$ in Eq. \eqref{A_Matrix}, must be negative.
	
	Figure \ref{stability} further illustrates how the stable region can be tuned by varying the optomechanical couplings $(G_\pm)$ [Fig. \ref{stability}(a)] and the phase angle $\theta$ [Fig. \ref{stability}(b)] associated with phonon hopping. In the absence of imperfect couplings, (i.e $G^m_\pm = 0$), the entire parameter regime satisfying $G_+ \leq G_-$
	remains stable, irrespective of the presence of phonon hopping.

	\section{ Mechanical Dark-mode Control for $J = 0 $}\label{dark_mode_J=0}

	Lets introduce the following mechanical bright ($B_+$) and dark ($B_-$) modes into the above Hamiltonian,
	\begin{equation*}
		\begin{cases}
			B_+ &= \frac{G_{01} b_1 + G_{02} b_2}{G_0} \\
			B_- &= \frac{G_{02} b_1 - G_{01} b_2}{G_0}
		\end{cases} \implies
		\begin{cases}
			& b_1 = \frac{A_+ G_{01} + A_- G_{02}}{G_0} \\
			& b_2 = \frac{A_+ G_{02} - A_- G_{01}}{G_0}
		\end{cases},
	\end{equation*}
	with $G_0 = \sqrt{G_{01}^2 + G_{02}^2}$. Using these expressions in the above Hamiltonian leads to the transformed Hamiltonian,

	\begin{align}
		H &= \sum_{j=\pm} \Delta_j B_j^\dagger B_j  + F_- (B_+^\dagger B_- + B_-^\dagger B_+), \\
		& - F_+ (a B_+^\dagger + B_+ a^\dagger) + \sum_{j=\pm} ( \tilde{F}_j a^\dagger B_j^\dagger + \tilde{F}_j B_j a)  +H_{diss} \nonumber
	\end{align}
	
	where we have defined $\Delta_{+(-)} =  \frac{\Omega}{|G_0|^2} \left[  |G_{01(02)}|^2 - |G_{02(01)}|^2 \right]$, $F_- = \frac{2 G_{01}G_{02}\Omega }{|G_0|^2}$, and $F_+ = G_0$, and $\tilde{F}_+ = \frac{G_{11}G_{01}+G_{12}G_{02}}{G_0}, \tilde{F}_- = \frac{G_{11}G_{02}-G_{12}G_{01}}{G_0}$

	\begin{figure}[t!]
		\includegraphics[width= 1 \linewidth]{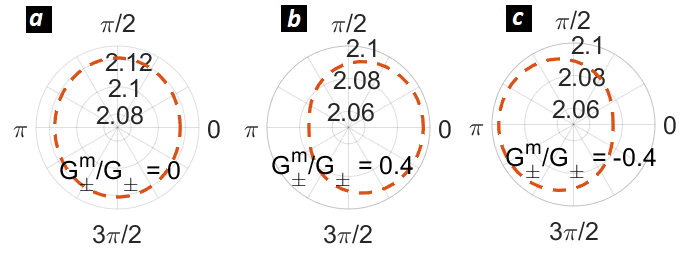}
		\caption{ $|B|_m$ in polar coordinates for different coupling imperfections (a) $G_\pm^m/G_\pm = 0$ (b) $G_\pm^m/G_\pm = 0.4$ and (c) $G_\pm^m/G_\pm = -0.4$. $J/\kappa= 1/150$ and all other parameters remain the same as in Fig. \ref{EN_NL}.
		}\label{polar_plot_ph}
	\end{figure}
	\section{Dark mode in the presence of coupling imperfection}\label{DM_coupling_imperfection}
	
	 Fig. \ref{polar_plot_ph} (a) shows the Bell nonlocality ($|B|_m$) to be maximized at $\theta=\pm\pi/2$ in case of $G^m_\pm = 0$. In case of $G^m_\pm \neq 0$, maximum of $|B|_m$ deviates from $\theta=\pm\pi/2$ (Fig. \ref{polar_plot_ph} (b and c)). The plot of $|B|_m$ in polar coordinate without and with the presence of coupling imperfection, moreover, follows Fig. \ref{nl_J}.

	\nocite{*}

	\bibliography{apssamp}
	\bibliographystyle{apsrev4-2}

\end{document}